# Harmonic Oscillator in Relativistic Minimal Length Quantum Mechanics

H. Hassanabadi[1*], S. Zarrinkamar[2] and E. Maghsoodi[1]

[1]Physics Department, Shahrood University of Technology, Shahrood, Iran

[2]Department of Basic Sciences, Garmsar Branch, Islamic Azad University, Garmsar, Iran

* h.hasanabadi@shahrodut.ac.ir

**Abstract**

We consider the Dirac equation with a generalized uncertainty principle in the presence of the Harmonic interaction and an external magnetic field. By doing the study the momentum space, the problem solved in an exact analytical manner and the eigenfucntions reported in terms of the hypergeometric functions.

## 1. Introduction

Quantum gravity, string theory, black-hole physics and doubly special relativity indicate that our ordinary Heisenberg uncertainty relation [1] should be modified into a generalized form which we call the generalized uncertainty principle (GUP) in the jargon [2,3]. Consequently, when the commutations relations, or equivalently the uncertainty principle, are modified, the wave equation is modified and thus the physics of the problem is altered as well. On the other hand, due to the experimental limitations and the energy scale, the theoretical studies become more urgent and appealing. There exist interesting papers that discuss various aspects of the problem with different wave equations of quantum mechanics including the Schrödinger and the relativistic wave equations [4-16]. Here, bearing in mind the impact of Dirac equation, Harmonic oscillator interaction and the GUP, we are going to consider the minimal length Dirac equation with a Harmonic term in the presence of an external magnetic field. On the contrary to ordinary Dirac equation, which has been solved by a lengthy list of phenomenological interaction [17-26], the equation has not been investigated in the minimal length formulation.

Our work is organized as follows. We first revisit the modified quantum mechanics due to minimal length. At the next step, we introduce two specific conditions of Dirac equation and, bearing in mind the physical significance of the harmonic oscillator interaction, we obtain solutions of the equation with this term when the effect an external magnetic field is included.

## 2. The Generalized Uncertainty Principle

In a proposed scenario, we consider the GUP

$$\Delta x \geq \frac{\hbar}{\Delta p} + \alpha l_p^2 \frac{\Delta p}{\hbar}, \qquad (1)$$

in which $\alpha$, called the the GUP parameter, is determined form a fundamental theory []. At energies much below the Planck mass, the second term in the right hand side of the latter vanishes and we recover the well-known Heisenberg uncertainty principle. The GUP of Eq. (1) is equivalent to the modified commutation relation [4-8]

$$[x_{op}, p_{op}] = i\hbar(1 + \beta p^2), \qquad 0 \leq \beta \leq 1 \qquad (2)$$

where $x_{op} = \hat{x}$ and $0 \leq \beta \leq 1$. When $\beta \to 0$, we recover the ordinary Heisenberg relation and the other extreme, i.e. $\beta \to 1$, corresponds to the so-called extreme quantum gravity. Eq. (2) gives the minimal length in this case as $(\Delta x)_{min} = 2l_p \sqrt{\alpha}$. A representation of $\hat{x}_i$ and $\hat{p}_i$ which satisfies Eq. (2) may be taken as [4-8]

$$\hat{x}_i = i\hbar(1 + \beta p^2)\frac{\partial}{\partial p_i}, \qquad \hat{p}_i = p_i. \qquad (3)$$

## 3. Dirac Equation with Mixed vector and scalar Harmonic potential

Dirac equation with scalar potential *S(r)* and vector potential *V (r)* in momentum space posses the form (in units $\hbar = c = 1$) [17]

$$\left[\vec{\alpha}.\vec{\Pi} + \beta(M + S(r))\right]\psi(\vec{p}) = \left[E - V(r)\right]\psi(\vec{p}), \qquad (4)$$

where E, $\vec{\Pi}$ and $M$ respectively denote the relativistic energy of the system, two-dimensional momentum operator and the mass of the fermionic particle. $\alpha$ and $\beta$ as usual are

$$\alpha = \begin{pmatrix} 0 & \vec{\sigma} \\ \vec{\sigma} & 0 \end{pmatrix}, \quad \beta = \begin{pmatrix} I & 0 \\ 0 & -I \end{pmatrix}, \qquad (5)$$

where $I$ is a $2 \times 2$ unitary matrix and the spin matrices are

$$\sigma_1 = \begin{pmatrix} 0 & 1 \\ 1 & 0 \end{pmatrix}, \sigma_2 = \begin{pmatrix} 0 & -i \\ i & 0 \end{pmatrix}. \qquad (6)$$

In the Pauli–Dirac representation, the wavefunction is represented as

$$\Psi(\vec{p}) = \begin{pmatrix} \varphi(\vec{p}) \\ \chi(\vec{p}) \end{pmatrix}, \qquad (7)$$

and, by a combination of Eqs. (4)-(7), we obtain the following coupled equations [17]

$$\sigma.\Pi \chi(\vec{p}) = [E - V(r) - M - S(r)]\varphi(\vec{p}), \qquad (8-a)$$
$$\sigma.\Pi \varphi(\vec{p}) = [E - V(r) + M + S(r)]\chi(\vec{p}). \qquad (8-b)$$

From now on, we use the famous notations

$$\Delta(r) = V(r) - S(r), \qquad (9-a)$$
$$\Sigma(r) = V(r) + S(r). \qquad (9-b)$$

Within the next sections, we will explore some special cases of the problem.

**2.1. The case of $\Delta(r) = 0$**

When the scalar potential $S(r)$ equals the vector potential $V(r)$, Eq. (9-a) becomes

$$\sigma.\Pi \chi^p_{n\lambda}(\vec{p}) = [E^p_{n\lambda} - M - 2V(r)]\varphi^p_{n\lambda}(\vec{p}), \qquad (10-a)$$

$$\chi^p_{n\lambda}(\vec{p}) = \frac{\sigma.\vec{\Pi}}{E^p_{n\lambda} + M} \varphi^p_{n\lambda}(\vec{p}). \qquad (10-b)$$

Substituting Eq. (10-b) into Eq. (10-a), we find

$$[\Pi^2 + 2(E^p_{n\lambda} + M)V(r)]\varphi^p_{n\lambda}(\vec{p}) = [E^{p\,2}_{n\lambda} - M^2]\varphi^p_{n\lambda}(\vec{p}). \qquad (11)$$

By considering $\vec{\Pi} = \vec{p} - \frac{e}{c}\vec{A}$, the Dirac equation for a charged particle moving in a constant magnetic field appears as [17]

$$[(\vec{p} - \frac{e}{c}\vec{A})^2 + 2(E^p_{n\lambda} + M)V(r)]\varphi^p_{n\lambda}(\vec{p}) = [E^{p\,2}_{n\lambda} - M^2]\varphi^p_{n\lambda}(\vec{p}), \qquad (12)$$

where the vector potential $\vec{A}$ is considered as $\vec{A} = (-\frac{B\hat{y}}{2}, \frac{B\hat{x}}{2}, 0)$ which corresponds to $\vec{B} = B\hat{z}$. For the case of potential, we work on a Harmonic potential
$$V(r) = V_0 r^2, \qquad (13)$$
and therefore the Eq. (12) takes the form

$$[(p_x - \frac{e}{c}A_x)\hat{i} + (p_y - \frac{e}{c}A_y)\hat{j}]^2 + 2(E^p_{n\lambda} + M)(\hat{x}^2 + \hat{y}^2) + M^2 - E^{p\,2}_{n\lambda}]\varphi^p_{n\lambda}(\vec{p}) = 0. \qquad (14)$$

in which

$$\hat{x} = i(1 + \beta p^2)\frac{\partial}{\partial p_x}, \qquad (15a)$$

$$\hat{y} = i(1 + \beta p^2)\frac{\partial}{\partial p_y}. \qquad (15b)$$

Writing $p_x = p\cos\vartheta$, $p_y = p\sin\vartheta$, Eq. (14) takes the form

$$\{(p_x + \frac{e}{c}\frac{B}{2}i(1+\beta p^2)\frac{\partial}{\partial p_y})(p_x + \frac{e}{c}\frac{B}{2}i(1+\beta p^2)\frac{\partial}{\partial p_y}) + (p_y - \frac{e}{c}\frac{B}{2}i(1+\beta p^2)\frac{\partial}{\partial p_x})$$

$$\times (p_y - \frac{e}{c}\frac{B}{2}i(1+\beta p^2)\frac{\partial}{\partial p_x})] + 2(E^p_{n\lambda} + M)W_0[(i(1+\beta p^2)\frac{\partial}{\partial p_x})(i(1+\beta p^2)\frac{\partial}{\partial p_x}) + (i(1+\beta p^2)\frac{\partial}{\partial p_y})$$

$$\times (i(1+\beta p^2)\frac{\partial}{\partial p_y})] + M^2 - E^{p\,2}_{n\lambda}\}\varphi^p_{n\lambda}(\vec{p}) = 0, \quad (16)$$

which, via

$$\frac{\partial}{\partial p_x} = \cos\vartheta\frac{\partial}{\partial p} - \frac{\sin\vartheta}{p}\frac{\partial}{\partial \vartheta},$$

$$\frac{\partial}{\partial p_y} = \sin\vartheta\frac{\partial}{\partial p} + \frac{\cos\vartheta}{p}\frac{\partial}{\partial \vartheta}, \quad (17)$$

can be alternatively written as

$$[-\frac{e^2B^2}{4c^2}(1+\beta p^2)^2 - 2(E^p_{n\lambda} + M)W_0(1+\beta p^2)^2]\frac{\partial^2}{\partial p^2} + [-\frac{e^2B^2}{2c^2}(1+\beta p^2)\beta p - \frac{e^2B^2}{4c^2}\frac{(1+\beta p^2)^2}{p}$$

$$-4(E^p_{n\lambda} + M)W_0(1+\beta p^2)\beta p - 2(E^p_{n\lambda} + M)W_0\frac{(1+\beta p^2)^2}{p}]\frac{\partial}{\partial p} + [-\frac{e^2B^2}{4c^2}\frac{(1+\beta p^2)^2}{p^2}$$

$$-2(E^p_{n\lambda} + M)W_0\frac{(1+\beta p^2)^2}{p^2}]\frac{\partial^2}{\partial \vartheta^2} + [\frac{eB}{2c}i(1+\beta p^2)]\frac{\partial}{\partial \vartheta} + p^2 + M^2 - E^{p\,2}_{n\lambda}]\varphi^p_{n\lambda}(\vec{p}) = 0. \quad (18)$$

Proposing the separation of the variables

$$\varphi^p_{n\lambda}(\vec{p}) = U^p_{n\lambda}(p)e^{i\lambda\vartheta}, \quad (19)$$

Eq. (18) transforms into

$$[-\frac{e^2B^2}{4c^2}(1+\beta p^2)^2 - 2(E^p_{n\lambda} + M)W_0(1+\beta p^2)^2]\frac{\partial^2}{\partial p^2} + [-\frac{e^2B^2}{2c^2}(1+\beta p^2)\beta p - \frac{e^2B^2}{4c^2}\frac{(1+\beta p^2)^2}{p}$$

$$-4(E^p_{n\lambda} + M)W_0(1+\beta p^2)\beta p - 2(E^p_{n\lambda} + M)W_0\frac{(1+\beta p^2)^2}{p}]\frac{\partial}{\partial p} + [p^2 - \frac{eB}{2c}(1+\beta p^2)\lambda$$

$$+ \frac{e^2B^2}{4c^2}\frac{(1+\beta p^2)^2}{p^2}\lambda^2 + 2\lambda^2(E^p_{n\lambda} + M)W_0\frac{(1+\beta p^2)^2}{p^2} + M^2 - E^{p\,2}_{n\lambda}]\varphi^p_{n\lambda}(\vec{p}) = 0. \quad (20)$$

To obtain solution of Eq. (20), we apply the transformation $z = -\beta p^2$ and rewrite the latter as

$$\frac{d^2 U^p_{n\lambda}(z)}{dz^2} + \frac{(1-\tau^p z)}{z(1-z)}\frac{dU^p_{n\lambda}(z)}{dz} + \frac{1}{z(1-z)}(\zeta^p z^2 + \nu^p z + \eta^p)U^p_{n\lambda}(z) = 0, \quad (21)$$

where

$$\mu^p = \frac{e^2 B^2}{c^2}\beta + 8\beta V_0 (E_{n\lambda}^p + M),$$

$$\tau^p = 2\frac{e^2 B^2}{\mu^p c^2}\beta + \frac{16\beta V_0}{\mu^p}(E_{n\lambda}^p + M).$$

$$\xi^p = -\frac{1}{\mu^p \beta} + \frac{eB}{2\mu^p c}\lambda - \frac{e^2 B^2}{4\mu^p c^2}\lambda^2 \beta - \frac{2V_0 \lambda^2 \beta}{\mu^p}(E_{n\lambda}^p + M),$$

$$v^p = -\frac{eB}{2\mu^p c}\lambda + \frac{e^2 B^2}{2\mu^p c^2}\lambda^2 \beta + \frac{4V_0 \lambda^2 \beta}{\mu^p}(E_{n\lambda}^p + M) + \frac{M^2}{\mu^p} - \frac{E_{n\lambda}^{p\,2}}{\mu^p},$$

$$\eta^p = -\frac{e^2 B^2}{4\mu^p c^2}\lambda^2 \beta - \frac{2V_0 \lambda^2 \beta}{\mu^p}(E_{n\lambda}^p + M). \qquad (22)$$

By comparing Eq. (21) with Eq. (*A*-6), we get

$$\alpha_1 = 1,\ \alpha_2 = \tau^p,\ \alpha_3 = 1,\ \xi_1 = -\zeta^p,\ \xi_2 = v^p,\ \xi_3 = -\eta^p,$$

$$\alpha_4 = 0,\ \alpha_5 = \frac{\tau^p}{2} - 1,\ \alpha_6 = (\frac{\tau^p}{2} - 1)^2 - \zeta^p,\ \alpha_7 = -v^p,$$

$$\alpha_8 = -\eta^p,\ \alpha_9 = (\frac{\tau^p}{2} - 1)^2 - \zeta^p - v^p - \eta^p,\ \alpha_{10} = 1 + 2\sqrt{-\eta^p},$$

$$\alpha_{11} = 2 + 2(\sqrt{(\frac{\tau^p}{2} - 1)^2 - \zeta^p - v^p - \eta^p} + \sqrt{-\eta^p}),\ \alpha_{12} = \sqrt{-\eta^p},$$

$$\alpha_{13} = \frac{\tau^p}{2} - 1 - (\sqrt{(\frac{\tau^p}{2} - 1)^2 - \zeta^p - v^p - \eta^p} + \sqrt{-\eta^p}), \qquad (23)$$

For this case, from Eq. (A-19), the wave function is

$$U_{n\lambda}^p(p) = (-\beta p^2)^{\sqrt{-\eta^p}}(1 + \beta p^2)^{-\frac{\tau^p}{2}+1+\sqrt{(\frac{\tau^p}{2}-1)^2 - \zeta^p - v^p - \eta^p}} P_n^{(2\sqrt{-\eta^p},\,2\sqrt{(\frac{\tau^p}{2}-1)^2 - \zeta^p - v^p - \eta^p})}(1 + 2\beta p^2), \qquad (24)$$

where $P_n^{(\alpha,\beta)}(1-2z)$ is the Jacobi polynomials or

$$P_n^{(\alpha,\beta)}(1-2z) = \frac{1}{n!} z^{-\alpha}(1-z)^{-\beta} \frac{d^n}{dz^n}[z^{n+\alpha}(1-z)^{n+\beta}], \qquad (25)$$

Therefore, $U_{n\lambda}^p(p)$ can be writing vs. the hypergeometric function as

$$U_{n\lambda}^p(p) = (-\beta p^2)^{\sqrt{-\eta^p}}(1 + \beta p^2)^{-\frac{\tau^p}{2}+1+\sqrt{(\frac{\tau^p}{2}-1)^2 - \zeta^p - v^p - \eta^p}}$$

$$\times {}_2F_1(-n, n + 2\sqrt{-\eta^p} + 2\sqrt{(\frac{\tau^p}{2}-1)^2 - \zeta^p - v^p - \eta^p} + 1; 2\sqrt{-\eta^p} + 1; -\beta p^2). \qquad (26)$$

which immediately gives

$$\varphi_{n\lambda}^{p}(\vec{p}) = (-\beta p^{2})^{\sqrt{-\eta^{p}}} (1+\beta p^{2})^{-\frac{\tau^{p}}{2}+1+\sqrt{(\frac{\tau^{p}}{2}-1)^{2}-\zeta^{p}-\nu^{p}-\eta^{p}}}$$

$$\times {}_{2}F_{1}(-n, n+2\sqrt{-\eta^{p}}+2\sqrt{(\frac{\tau^{p}}{2}-1)^{2}-\zeta^{p}-\nu^{p}-\eta^{p}}+1; 2\sqrt{-\eta^{p}}+1; -\beta p^{2})e^{i\lambda\vartheta}, \qquad (27)$$

and

$$\Psi_{n\lambda}^{p}(\vec{p}) = N_{n\lambda}\begin{pmatrix} 1 \\ \dfrac{\sigma.\Pi}{E+M} \end{pmatrix}(-\beta p^{2})^{\sqrt{-\eta^{p}}}(1+\beta p^{2})^{-\frac{\tau^{p}}{2}+1+\sqrt{(\frac{\tau^{p}}{2}-1)^{2}-\zeta^{p}-\nu^{p}-\eta^{p}}}$$

$$\times {}_{2}F_{1}(-n, n+2\sqrt{-\eta^{p}}+2\sqrt{(\frac{\tau^{p}}{2}-1)^{2}-\zeta^{p}-\nu^{p}-\eta^{p}}+1; 2\sqrt{-\eta^{p}}+1; -\beta p^{2})e^{i\lambda\vartheta}, \qquad (28)$$

where $N_{n\lambda}$ is the normalization constant and from Eq. (A-15), we have

$$\tau^{p} n - (2n+1)(\frac{\tau^{p}}{2}-1) + (2n+1)(\sqrt{(\frac{\tau^{p}}{2}-1)^{2}-\zeta^{p}-\nu^{p}-\eta^{p}}+\sqrt{-\eta^{p}}) + n(n-1) - \nu^{p}$$

$$-2\eta^{p} + 2\sqrt{-\eta^{p}((\frac{\tau^{p}}{2}-1)^{2}-\zeta^{p}-\nu^{p}-\eta^{p})} = 0. \qquad (29)$$

## 2.2. The case of $\Sigma(r) = 0$

In this section, the scalar potential $S(r)$ is equal to the vector potential $V(r)$ and Eq. (8) becomes

$$\sigma.\Pi \varphi_{n\lambda}^{s}(\vec{p}) = [E_{n\lambda}^{s} + M - 2V(r)]\chi_{n\lambda}^{s}(\vec{p}). \qquad (30-b)$$

$$\varphi_{n\lambda}^{s}(\vec{p}) = \frac{\sigma.\vec{\Pi}}{E-M}\chi_{n\lambda}^{s}(\vec{p}). \qquad (30-b)$$

which, after eliminating one component in favor of the other, gives the decoupled equation

$$[\Pi^{2} + 2(E_{n\lambda}^{s} - M)V(r)]\chi_{n\lambda}^{s}(\vec{p}) = [E_{n\lambda}^{s\,2} - M^{2}]\chi_{n\lambda}^{s}(\vec{p}). \qquad (31)$$

Therefore, the equation with a constant magnetic field has the form

$$[(\vec{p} - \frac{e}{c}\vec{A})^{2} + 2(E_{n\lambda}^{s} - M)V(r)]\chi_{n\lambda}^{s}(\vec{p}) = [E_{n\lambda}^{s\,2} - M^{2}]\chi_{n\lambda}^{s}(\vec{p}). \qquad (32)$$

which, after inserting the Harmonic potential, appears as

$$[(p_{x} - \frac{e}{c}A_{x})\hat{i} + (p_{y} - \frac{e}{c}A_{y})\hat{j}]^{2} + 2(E_{n\lambda}^{s} - M)(\hat{x}^{2} + \hat{y}^{2}) + M^{2} - E_{n\lambda}^{s\,2}]\chi_{n\lambda}^{s}(\vec{p}) = 0, \qquad (33)$$

or equivalently

$$\{(p_x + \frac{e}{c}\frac{B}{2}i(1+\beta p^2)(\sin\vartheta\frac{\partial}{\partial p} + \frac{\cos\vartheta}{p}\frac{\partial}{\partial\vartheta}))(p_x + \frac{e}{c}\frac{B}{2}i(1+\beta p^2)(\sin\vartheta\frac{\partial}{\partial p} + \frac{\cos\vartheta}{p}\frac{\partial}{\partial\vartheta}))$$

$$+(p_y - \frac{e}{c}\frac{B}{2}i(1+\beta p^2)(\cos\vartheta\frac{\partial}{\partial p} - \frac{\sin\vartheta}{p}\frac{\partial}{\partial\vartheta}))\times(p_y - \frac{e}{c}\frac{B}{2}i(1+\beta p^2)(\cos\vartheta\frac{\partial}{\partial p} - \frac{\sin\vartheta}{p}\frac{\partial}{\partial\vartheta}))]$$

$$+2(E_{n\lambda}^p - M)V_0[(i(1+\beta p^2)(\cos\vartheta\frac{\partial}{\partial p} - \frac{\sin\vartheta}{p}\frac{\partial}{\partial\vartheta}))(i(1+\beta p^2)(\cos\vartheta\frac{\partial}{\partial p} - \frac{\sin\vartheta}{p}\frac{\partial}{\partial\vartheta}))$$

$$+(i(1+\beta p^2)(\sin\vartheta\frac{\partial}{\partial p} + \frac{\cos\vartheta}{p}\frac{\partial}{\partial\vartheta}))(i(1+\beta p^2)(\sin\vartheta\frac{\partial}{\partial p} + \frac{\cos\vartheta}{p}\frac{\partial}{\partial\vartheta}))] + M^2 - E_{n\lambda}^{p\,2}\}\chi_{n\lambda}^s(\vec{p}) = 0. \quad (34)$$

Eq. (34) is more compactly is written in the form

$$[-\frac{e^2B^2}{4c^2}(1+\beta p^2)^2 - 2(E_{n\lambda}^p - M)V_0(1+\beta p^2)^2]\frac{\partial^2}{\partial p^2} + [-\frac{e^2B^2}{2c^2}(1+\beta p^2)\beta p - \frac{e^2B^2}{4c^2}\frac{(1+\beta p^2)^2}{p}$$

$$-4(E_{n\lambda}^p - M)V_0(1+\beta p^2)\beta p - 2(E_{n\lambda}^p - M)V_0\frac{(1+\beta p^2)^2}{p}]\frac{\partial}{\partial p} + [-\frac{e^2B^2}{4c^2}\frac{(1+\beta p^2)^2}{p^2}$$

$$-2(E_{n\lambda}^p - M)V_0\frac{(1+\beta p^2)^2}{p^2}]\frac{\partial^2}{\partial\vartheta^2} + [\frac{eB}{2c}i(1+\beta p^2)]\frac{\partial}{\partial\vartheta} + p^2 + M^2 - E_{n\lambda}^{p\,2}]\chi_{n\lambda}^s(\vec{p}) = 0, \quad (35)$$

For proceeding further, via the solution

$$\chi_{n\lambda}^s(\vec{p}) = U_{n\lambda}^s(p)e^{i\lambda\vartheta}, \quad (36)$$

we remove the angle dependence and obtain

$$[-\frac{e^2B^2}{4c^2}(1+\beta p^2)^2 - 2(E_{n\lambda}^p - M)V_0(1+\beta p^2)^2]\frac{\partial^2}{\partial p^2} + [-\frac{e^2B^2}{2c^2}(1+\beta p^2)\beta p - \frac{e^2B^2}{4c^2}\frac{(1+\beta p^2)^2}{p}$$

$$-4(E_{n\lambda}^p - M)V_0(1+\beta p^2)\beta p - 2(E_{n\lambda}^p - M)V_0\frac{(1+\beta p^2)^2}{p}]\frac{\partial}{\partial p} + [p^2 - \frac{eB}{2c}(1+\beta p^2)\lambda$$

$$+\frac{e^2B^2}{4c^2}\frac{(1+\beta p^2)^2}{p^2}\lambda^2 + 2\lambda^2(E_{n\lambda}^p - M)V_0\frac{(1+\beta p^2)^2}{p^2} + M^2 - E_{n\lambda}^{p\,2}]\chi_{n\lambda}^s(\vec{p}) = 0. \quad (37)$$

The transformation $z = -\beta p^2$ brings the latter into the form

$$\frac{d^2U_{n\lambda}^s(z)}{dz^2} + \frac{(1-\tau^s z)}{z(1-z)}\frac{dU_{n\lambda}^s(z)}{dz} + \frac{1}{z(1-z)}(\zeta^s z^2 + \nu^s z + \eta^s)U_{n\lambda}^s(z) = 0, \quad (38)$$

with

$$\mu^s = \frac{e^2 B^2}{c^2}\beta + 8\beta V_0 (E^s_{n\lambda} - M),$$

$$\tau^s = 2\frac{e^2 B^2}{\mu^s c^2}\beta + \frac{16\beta V_0}{\mu^s}(E^s_{n\lambda} - M).$$

$$\xi^s = -\frac{1}{\mu^s \beta} + \frac{eB}{2\mu^s c}\lambda - \frac{e^2 B^2}{4\mu^s c^2}\lambda^2 \beta - \frac{2V_0 \lambda^2 \beta}{\mu^s}(E^s_{n\lambda} - M),$$

$$\nu^s = -\frac{eB}{2\mu^s c}\lambda + \frac{e^2 B^2}{2\mu^s c^2}\lambda^2 \beta + \frac{4V_0 \lambda^2 \beta}{\mu^s}(E^s_{n\lambda} - M) + \frac{M^2}{\mu^s} - \frac{E^{s\,2}_{n\lambda}}{\mu^s},$$

$$\eta^s = -\frac{e^2 B^2}{4\mu^s c^2}\lambda^2 \beta - \frac{2V_0 \lambda^2 \beta}{\mu^s}(E^s_{n\lambda} - M). \qquad (39)$$

Finally, the corresponding eigenfunctions are

$$U^s_{n\lambda}(p) = (-\beta p^2)^{\sqrt{-\eta^s}}(1+\beta p^2)^{-\frac{\tau^s}{2}+1+\sqrt{(\frac{\tau^s}{2}-1)^2 - \zeta^s - \nu^s - \eta^s}}$$

$$\times {}_2F_1(-n, n + 2\sqrt{-\eta^s} + 2\sqrt{(\frac{\tau^s}{2}-1)^2 - \zeta^s - \nu^s - \eta^s} + 1; 2\sqrt{-\eta^s} + 1; -\beta p^2). \quad (40)$$

which immediately gives

$$\chi^s_{n\lambda}(\vec{p}) = (-\beta p^2)^{\sqrt{-\eta^s}}(1+\beta p^2)^{-\frac{\tau^s}{2}+1+\sqrt{(\frac{\tau^s}{2}-1)^2 - \zeta^s - \nu^s - \eta^s}}$$

$$\times {}_2F_1(-n, n + 2\sqrt{-\eta^s} + 2\sqrt{(\frac{\tau^s}{2}-1)^2 - \zeta^s - \nu^s - \eta^s} + 1; 2\sqrt{-\eta^s} + 1; -\beta p^2)e^{i\lambda\vartheta}, \qquad (41)$$

and

$$\Psi^s_{n\lambda}(\vec{p}) = N_{n\lambda}\begin{pmatrix} 1 \\ \dfrac{\sigma.\Pi}{E-M} \end{pmatrix}(-\beta p^2)^{\sqrt{-\eta^s}}(1+\beta p^2)^{-\frac{\tau^s}{2}+1+\sqrt{(\frac{\tau^s}{2}-1)^2 - \zeta^s - \nu^s - \eta^s}}$$

$$\times {}_2F_1(-n, n + 2\sqrt{-\eta^s} + 2\sqrt{(\frac{\tau^s}{2}-1)^2 - \zeta^s - \nu^s - \eta^s} + 1; 2\sqrt{-\eta^s} + 1; -\beta p^2)e^{i\lambda\vartheta}, \quad (42)$$

where $N_{n\lambda}$ is the normalization constant and f with the corresponding energy relation being determined from

$$\tau^s n - (2n+1)(\frac{\tau^s}{2}-1) + (2n+1)(\sqrt{(\frac{\tau^s}{2}-1)^2 - \zeta^s - \nu^s - \eta^s} + \sqrt{-\eta^s}) + n(n-1) - \nu^s$$

$$-2\eta^s + 2\sqrt{-\eta^s((\frac{\tau^s}{2}-1)^2 - \zeta^s - \nu^s - \eta^s)} = 0 \quad (43)$$

## 3. Conclusion

By considering the generalized quantum mechanics, exact solution of the two-dimensional Dirac equation in a magnetic field in the presence of a minimal length has been obtained in terms of hypergeometric functions. We obtained a second-order differential equation which resembles the standard Schrodinger equation. We also reported the eigenfunctions and the energy eigenvalues.

## Appendix

**Parametric formulation of Nikiforov-Uvarov method**

The NU method, in its parametric form, solves [27]

$$\frac{d^2}{ds^2}\psi_n(s) + \frac{\alpha_1 - \alpha_2 s}{s(1-\alpha_3 s)}\frac{d}{ds}\psi_n(s) + \frac{-\xi_1 s^2 + \xi_2 s - \xi_3}{[s(1-\alpha_3 s)]^2}\psi_n(s) = 0 \qquad (A-1)$$

Here, we give only the basic ingredients of the generalized NU method. In this case, we can obtain [27]

$$\tilde{\tau}(s) = \alpha_1 - \alpha_2 s, \qquad (A-2)$$
$$\sigma(s) = s(1-\alpha_3 s), \qquad (A-3)$$
$$\tilde{\sigma}(s) = -\xi_1 s^2 + \xi_2 s - \xi_3, \qquad (A-4)$$

Inserting the above equations into Eq. (*A*-6) leads to [27]

$$\pi(s) = \alpha_4 + \alpha_5 s \pm \sqrt{(\alpha_6 - k\alpha_3)s^2 + (\alpha_7 + k)s + \alpha_8}, \qquad (A-5)$$

where

$$\alpha_4 = \frac{1}{2}(1-\alpha_1), \qquad (A-6)$$
$$\alpha_5 = \frac{1}{2}(\alpha_2 - 2\alpha_3), \qquad (A-7)$$
$$\alpha_6 = \alpha_5^2 + \xi_1, \qquad (A-8)$$
$$\alpha_7 = 2\alpha_4\alpha_5 - \xi_2, \qquad (A-9)$$
$$\alpha_8 = \alpha_4^2 + \xi_3. \qquad (A-10)$$

and

$$\alpha_9 = \alpha_3\alpha_7 + \alpha_3^2\alpha_8 + \alpha_6, \qquad (A.11)$$

one can easily see that different *k* values lead to different $\pi(s)$s. If we take

$$k = -(\alpha_7 + 2\alpha_3\alpha_8) - 2\sqrt{\alpha_8\alpha_9}, \qquad (A.12)$$

$\pi(s)$ becomes

$$\pi(s) = \alpha_4 + \alpha_5 s - [(\sqrt{\alpha_9} + \alpha_3\sqrt{\alpha_8})s - \sqrt{\alpha_8}), \qquad (A.13)$$

and then we find [27]

$$\tau(s) = \alpha_1 + 2\alpha_4 - (\alpha_2 - 2\alpha_5)s - [(\sqrt{\alpha_9} + \alpha_3\sqrt{\alpha_8})s - \sqrt{\alpha_8}]. \qquad (A.14)$$

with the above results as follows [27]

$$\alpha_2 n - (2n+1)\alpha_5 + (2n+1)(\sqrt{\alpha_9} + \alpha_3\sqrt{\alpha_8}) + n(n-1)\alpha_3 + \alpha_7 + 2\alpha_3\alpha_8 + 2\sqrt{\alpha_8\alpha_9} = 0 \qquad (A.15)$$

In order to obtain the wave functions, one can use the following relations [27]

$$\rho(s) = s^{\alpha_{10}-1}(1-\alpha_3 s)^{(\alpha_{11}/\alpha_3)-\alpha_{10}-1}, \qquad (A.16)$$

$$\Phi_n(s) = P_n^{(\alpha_{10}-1,(\alpha_{11}/\alpha_3)-\alpha_{10}-1)}(1-2\alpha_3 s), \qquad (A.17)$$

$$W(s) = s^{\alpha_{12}}(1-\alpha_3 s)^{-\alpha_{12}-(\alpha_{13}/\alpha_3)}, \qquad (A.18)$$

$$\Psi_n(s) = s^{\alpha_{12}}(1-\alpha_3 s)^{-\alpha_{12}-(\alpha_{13}/\alpha_3)} P_n^{(\alpha_{10}-1,(\alpha_{11}/\alpha_3)-\alpha_{10}-1)}(1-2\alpha_3 s), \qquad (A.19)$$

where $P_n^{(\alpha_{10}-1,(\alpha_{11}/\alpha_3)-\alpha_{10}-1)}(1-2\alpha_3 s)$ is the Jacobi polynomials and [27]

$$\alpha_{10} = \alpha_1 + 2\alpha_4 + 2\sqrt{\alpha_8}, \qquad (A.20)$$

$$\alpha_{11} = \alpha_2 - 2\alpha_5 + 2(\sqrt{\alpha_9} + \alpha_3\sqrt{\alpha_8}), \qquad (A.21)$$

$$\alpha_{12} = \alpha_4 + \sqrt{\alpha_8}, \qquad (A.22)$$

$$\alpha_{13} = \alpha_5 - (\sqrt{\alpha_9} + \alpha_3\sqrt{\alpha_8}). \qquad (A.23)$$

**References**

[1] H. Kragh, "Arthur March,Werner Heisenberg, and the search for a smallest length," Revue d'Histoire des Sciences, vol. 8, no. 4, pp. 401–434 (1995).

[2]D. J. Gross and P. F. Mende, Nucl. Phys. B 303, 407 (1988).

[3]D. Amati, M. Ciafaloni, and G. Veneziano, Phys. Lett B 216, 41 (1989).

[4]D. J. Gross and P. F. Mende, Nucl. Phys. B 303, 407 (1988).

[5] D. Amati, M. Ciafaloni, and G. Veneziano, Phys. Lett B 216, 41 (1989).

[6] M. Maggiore, Phys. Lett. B 319, 83 (1993).8

[7] G. Amelino-Camelia, J. Ellis, N. E. Mavromatos, and D.V. Nanopoulos, Mod. Phys. Lett. A 12, 2029 (1997).

[8] F. Brau, F. Buisseret, Phys. Rev. D 74: 036002, (2006).

[9] H. Hassanabadi, S. Zarrinkamar, E. Maghsoodi, Phys. Lett. B, 718, 678 (2012).

[10] H. Hassanabadi, Z. Molaee, S. Zarrinkamar, Eur. Phys. J. C (2012)72:2217.

[11] L. Menculini, O. Panella, and P. Roy, Phys. Rev. D **87**, 065017 (2013).

[12] B. Mirza and M. Mohadesi, Commun. Theor. Phys. 42, 664 (2004).

[13] I. Dadic, L. Jonke, S. Meljanac, Phys.Rev. D67 087701 (2003).

[14] S. Benczik, L. N. Chang, D. Minic, N. Okamura, S. Rayyan, T. Takeuchi, Phys. Rev. D 66, 026003 (2002).

[15] L. N. Chang, Z. Lewis, D. Minic and T. Takeuchi, AdHEP Vol. 2011, Article ID 493514, 30 pages, doi:10.1155/2011/493514.

[16] K. Konishi, G. Paffuti, P. Provero, Phys. Lett. B234, 276 (1990).

[17] J. N. Ginocchio, Phys. Rep. 414 (2005) 165.

[18] C. Berkdemir, A. Berkdemir and J. Han, Chem. Phys. Lett. 417, 326 (2006) 326.

[19] S. H. Dong, G. H. Sun and D. Popov, J. Math. Phys. 44(10), 4467 (2003).

[20] S. H. Dong and Z. Q. Ma, Found. Phys. Lett.**15** (2),171 (2002).

[21] X. Y. Gu,Z. Q. Ma and S. H. Dong, Phys. Rev. A **67**(6),062715(2003).

[22] W. C. Qiang and S. H. Dong, Journal of Physics A 39, 8663 (2006).

[23] G. F. Wei and S. H. Dong, Phys. Lett. B 686, 288 (2010).

[24] G. F. Wei and S. H. Dong, EPJA 46,207 (2010).

[25] G. F. Wei, C.-Y. Long S.-H. Dong, Phys. Lett. A 372, 2592 (2008).

[26] G. H. Sun and S. H. Dong, Mod. Phys. Lett. A 25, 2849 (2010).

[27] C. Tezcan and R. Sever, Int. J. Theor. Phys. 48 (2009) 337.